\documentclass[a4paper,11pt]{article}
\pdfoutput=1 % if your are submitting a pdflatex (i.e. if you have
\usepackage{jcappub} % for details on the use of the package, please
\usepackage{xcolor}
\usepackage[T1]{fontenc} % if needed
\usepackage{bm}
\usepackage{epsfig}
\usepackage[lofdepth,lotdepth,caption=false]{subfig}
\usepackage{graphicx}% Include figure files

\usepackage{float}
\usepackage{amsmath,amssymb}
\usepackage{textcomp}
\usepackage{gensymb}
\usepackage[utf8]{inputenc}
\usepackage{setspace}
\usepackage{hyperref}
\usepackage[normalem]{ulem}
\usepackage{pdflscape}
\usepackage{multirow}
\usepackage{fancyvrb}
\def\beq{\begin{equation}}
\def\enq{\end{equation}}
\def\bea{\begin{eqnarray}}
\def\ena{\end{eqnarray}}

\begin{document}

\title{Evaporating Primordial Black Holes in Gamma Ray and Neutrino Telescopes}
\author{Antonio Capanema,}
\author{AmirFarzan Esmaeili,}
\author{Arman Esmaili}
\emailAdd{antoniogalvao@aluno.puc-rio.br}
\emailAdd{a.farzan.1993@aluno.puc-rio.br}
\emailAdd{arman@puc-rio.br}
\affiliation{Departamento de F\'isica, Pontif\'icia Universidade Cat\'olica do Rio de Janeiro,\\
Rua Marquês de São Vicente 225, Rio de Janeiro, Brazil}

\abstract{A primordial black hole in the last stages of evaporation and located in the local neighborhood can produce a detectable signal in gamma ray and neutrino telescopes. We re-evaluate the expected gamma ray and neutrino fluxes from these transient point events and discuss the consequences for existing constraints. For gamma rays we improve the current bounds by a factor of few, while for neutrinos we obtain significantly different results than the existing literature. The capability and advantages of neutrino telescopes in the search for primordial black holes is discussed thoroughly. The correlations of gamma ray and neutrino energy and time profiles will be promoted as a powerful tool in identifying the primordial black holes, in case of detection.}
\maketitle
\date{\today}

%%%%%%%%%%%%%%%%%%%%%%%%%%%%%%%%%
%%%%%%%%%%%%%%%%%%%%%%%%%%%%%%%%%
\section{\label{sec:intro}Introduction}
%%%%%%%%%%%%%%%%%%%%%%%%%%%%%%%%%
%%%%%%%%%%%%%%%%%%%%%%%%%%%%%%%%%

One of the prominent predictions of General Relativity (GR) is the existence of black hole (BH), a deformed spacetime around a sufficiently concentrated mass where all geodesics terminate at a singularity point (or region of zero volume). As far as GR is concerned, there is no restriction on the BH's mass; any mass $M$ distributed inside the Schwarzschild radius $r_s=2M$ will form a BH (throughout this paper, we use the natural units where $\hbar = c = k_b = G = 1$). However, the formation mechanism of a BH strongly restricts its possible masses. The gravitational collapse of astrophysical objects (stars) can lead to BHs with masses only between $\sim  2M_\odot$~\cite{Kalogera:1996tto} and $\sim 50M_\odot$~\cite{Leung:2019vdy}, which has been observed in X-ray binaries~\cite{Casares:2013tpa,10.1093/mnras/271.1.L10}. Merger events can create heavier BHs, potentially up to a few hundreds of solar masses, whose existence has been established observationally via gravitational waves~\cite{Abbott:2016blz}. Extremely heavier BHs of up to $\sim10^9M_\odot$ have been detected ubiquitously at the centers of galaxies (including ours), although their formation mechanism is still not well understood.

Meanwhile, on the $\lesssim M_\odot$ part of the mass spectrum, both the formation and the detection of BHs are challenging. In a seminal paper~\cite{Zel'dovich:1967dhs}, Zel’dovich and Novikov suggested the formation of such BHs via the gravitational collapse of overdense regions in the early universe, which have been fittingly named primordial black holes (PBHs). The formation rate of PBHs depends on the details of the initial power spectrum of the early universe, and the required conditions for having a relevant population of PBHs are still under debate (see \cite{Villanueva-Domingo:2021gjh}). Inspection of the early universe density fluctuations reveals that PBHs with masses $\sim 5M_\odot (t/10^{-4}~{\rm s})$ indeed can be formed at a time $t$ after the Big Bang~\cite{Carr:1975qj}. The thermodynamics of BHs plays an important role over cosmological timescales for light PBHs. Hawking showed in \cite{Hawking:1974sod,Hawking:1975zls} that a BH of mass $M$ loses its mass in the form of radiated particles with a quasi-black-body spectrum at temperature $T=(8\pi M)^{-1}$. This so-called Hawking radiation is only appreciable for light BHs (it is extremely small for stellar-mass BHs), and its very existence leads to a complete evaporation of PBHs with masses $\lesssim {\rm few}\times 10^{14}$~g within the age of the Universe\footnote{Strong limits on the abundance of PBHs with masses $\lesssim 10^{14}$~g can be derived from BBN considerations~\cite{Carr:2010mbp}, early neutrino and photon production~\cite{Bugaev:2002cmx,Bugaev:2009rmv}, reionization~\cite{He:2002chs,Mack:2008swa}, and even contribution to the dark matter if stable Planck-mass remnants are left behind from their evaporation~\cite{MacGibbon:1987sie,Barrow:1992eqb,Chen:2003rit}.}. PBHs with mass $\sim 10^{15}$~g, which (if they exist) have been produced at $t\sim10^{-23}$~s, are expected to be breathing their last breaths at the present time and are the subject of this paper\footnote{PBHs with masses $\gtrsim10^{15}$~g would still be alive and may account for a considerable fraction of the dark matter \cite{Chapline:1975ned} (see \cite{Villanueva-Domingo:2021gjh,Green:2020rgp,Carr:2020qwj} for recent reviews).}.

Hawking radiation leads to a runaway mass loss process: as the BH's mass decreases, the rate of energy loss increases dramatically ($\Dot{E} \sim M^{-2}$~\cite{Page:1976cco}) until the BH practically vanishes after emitting all particles in the Standard Model (and any other hypothetical beyond the Standard Model particle) up to Planck scale energies. In the last few hundreds of seconds of the BH's life, when its temperature exceeds the quarks' and gauge bosons' masses, the spectrum of radiated particles significantly deviates from a greybody spectrum due to secondary emission from hadronization and electroweak corrections. Among the stable final products of PBH evaporation, gamma rays have been most extensively used by the H.E.S.S.~\cite{Tavernier:2019dss,Glicenstein:2013dps}, HAWC~\cite{Albert:2020doa}, Milagro~\cite{Abdo:2015scw}, VERITAS~\cite{Archambault:2017tos} and Fermi-LAT~\cite{Ackermann:2018sps} experiments in the search for local (within $\sim$ a parsec) PBHs, of course, all with null results. These searches provide independent and complementary bounds to gamma ray observations at larger distance scales, e.g. Galactic scales using EGRET data~\cite{Wright:1996ncw,Lehuocq:2009ugr} (refined in~\cite{Carr:2010mbp,Carr:2016vur}) and cosmological scales using the extragalactic gamma ray background (EGB)~\cite{Page:1976gpe,Carr:2010mbp}. Besides gamma rays, other particles such as antiprotons~\cite{Abe:2012dpv}, electrons/positrons~\cite{Laha:2019ers,Boudaud:2019vrx,Dasgupta:2020cnq} and neutrinos \cite{Dasgupta:2020cnq} have also been used to constrain the density of $(10^{15}-10^{17})$~g PBHs and their fractional contribution to dark matter, $f_{\rm PBH}$. Strong upper limits on $f_{\rm PBH}$ for low-mass PBHs have also been obtained from EGB~\cite{Arbey:2020vnr}, cosmic microwave background anisotropy~\cite{Clark:2017whe}, Galactic gamma ray~\cite{Coogan:2021fnp,Laha:2020cqn}, and 21-cm~\cite{Mittal:2021gor} measurements, the latter yielding the strongest bounds to date: $f_{\rm PBH} \lesssim 10^{-9.7}$ at 95\% C.L. for PBHs with mass $\sim10^{15}$~g. Future X- and gamma ray experiments are expected to tighten these bounds even further, especially for rotating PBHs \cite{Ghosh:2021cej,Ray:2021cen}. One may ask whether, in light of these strong bounds, searches for PBHs in the local neighborhood is viable. Considering the local dark matter density $0.0133~M_\odot$~pc$^{-3}$~\cite{Guo:2020noi} and assuming a uniform PBH distribution throughout the universe, the bound on $f_{\rm PBH}$ leads to an average distance between two PBH bursts no shorter than $\sim 6\times 10^{-3}$~pc. Also, this distance scale can be significantly shorter in case of any (expected) clustering in the PBH's distribution. Since these distance scales are within the reach of gamma ray and neutrino observatories, direct observation of local PBH bursts keeps being motivated.

The idea of using neutrinos in the search for evaporating PBHs has been introduced in~\cite{Halzen:1995skd} (see~\cite{Carr:1976dbw} for the early idea of using stable particles is the search for PBHs), by taking into account the emission of secondary neutrinos and in view of upcoming neutrino telescopes, which are now a reality. In this paper we elaborate on this possibility and explore the extent of its feasibility in the current and future neutrino telescopes, as well as its synergy with gamma ray experiments in a multimessenger approach. As for the expected neutrino production by an evaporating PBH, with respect to previous works~\cite{Halzen:1995skd,Dave:2019fus}, we improve on the calculation by using the \verb+BlackHawk+ code~\cite{Arbey:2019had} for the primary particle emission and incorporate the \verb+HDMSpectra+ code~\cite{Bauer:2021vjs}, which is the state-of-the-art code for secondary computation. Applying these improvements to the PBH gamma ray production shows qualitative and quantitative differences to previous works which enables us to improve the existing limits from gamma ray experiments. We also discuss the three-fold advantage of using neutrinos in the search for evaporating PBHs: a) Neutrinos would be able to reach the earth independent of the existence of a speculated optically thick photosphere \cite{Heckler:1997bmo,Heckler:1997gry,MacGibbon:2008gfr} around PBHs which absorbs the gamma rays. The same applies for PBHs embedded in high density and/or magnetic field regions, where the outgoing flux of all the particles except for neutrinos will be suppressed. b) Extensive air shower experiments for gamma ray detection have a limited field of view -- $\sim2$~sr ($\sim 16\%$ of the sky) for HAWC~\cite{Abeysekara:2012ffp} and $\sim1$~sr for LHAASO~\cite{Bai:2019khm} -- while neutrino telescopes cover at least $2\pi$~sr of the sky with almost zero dead time of the detector. c) In case we observe a flare of neutrinos or gamma rays, pinpointing its origin is crucial. An orphan gamma ray flare can be attributed to a conventional leptonic process. An orphan neutrino flare matches the expectation for a gamma-ray-opaque source, as is favored by multi-messenger analyses~\cite{Capanema:2020rjj,Capanema:2021puc}. The simultaneous observation of neutrino and gamma ray flares, with the peculiar energy and time profile correlations of an evaporating PBH, can easily exclude the former alternatives.

This paper is structured as follows: in section~\ref{sec:rad}, we lay out the computational framework of the neutrino and gamma ray radiation from an evaporating PBH, comparing our updated emission spectra from \verb+BlackHawk+ and \verb+HDMSpectra+ to previous results in the literature. In section \ref{sec:detect}, we estimate IceCube's sensitivity to PBH bursts and set the corresponding upper limit on their local rate density from the 10-years neutrino data set. Section~\ref{sec:gamma-nu} is devoted to the multimessenger $\gamma/\nu$ approach in identifying a PBH event. Discussions and conclusions are summarized in section~\ref{sec:concl}.

%%%%%%%%%%%%%%%%%%%%%%%%%%%%%%%%%
%%%%%%%%%%%%%%%%%%%%%%%%%%%%%%%%%
\section{\label{sec:rad}Neutrinos and gamma rays from Hawking radiation}
%%%%%%%%%%%%%%%%%%%%%%%%%%%%%%%%%
%%%%%%%%%%%%%%%%%%%%%%%%%%%%%%%%%

The no-hair theorem \cite{Israel:1967fdk,Israel:1968did,Carter:1971spd,Hawking:1972vck,Robinson:1975jvc,Mazur:1982eps} states that a BH, for an external observer, can be fully characterised by specifying its mass, electric charge and angular momentum. However, BH evaporation causes it to discharge~\cite{Zaumen:1974aks,Carter:1974tus,Gibbons:1975rdp} and lose angular momentum~\cite{Page:1976ycs} at a much faster rate than it loses mass, once the BH approaches the end of its life. This allows us to safely assume that PBHs which are currently on their final hours are simply Schwarzschild BHs and parameterized only by their mass $M$. Nevertheless, at the end of this section we briefly comment on the rotating PBHs.

In terms of $M$, the temperature of a BH is $T = (8\pi M)^{-1}$ \cite{Hawking:1974sod,Hawking:1975zls} and its mass loss rate due to Hawking radiation can be obtained by solving the following equation of energy conservation~\cite{Page:1976cco}
\begin{equation}\label{eq:mass-time}
    \frac{{\rm d}M}{{\rm d}t} = -\frac{\alpha(M)}{M^2}~,
\end{equation}
where $\alpha(M)$ takes into account the total energy carried away by the Hawking radiation and can be calculated by summing over all particle species $i$ of the Standard Model (quarks, charged leptons, neutrinos, gauge bosons and the Higgs)\footnote{To a very good approximation, the emission of particle $i$ becomes non-negligible once the BH temperature approaches its rest mass \cite{MacGibbon:1990duy}.}
\begin{equation}
\label{eq:alpham}
    \alpha(M) = M^2 \sum_i \int_0^\infty {\rm d}E\, E\,\frac{{\rm d}^2 N^i_{\rm p}}{{\rm d}t {\rm d}E}(M,E)~.
\end{equation}
The graph of $\alpha(M)$ as function of $M$ can be found in Fig. 2 of~\cite{Baker:2021bhr}. In Eq.~(\ref{eq:alpham}), ${\rm d}^2 N^i_{\rm p}/{\rm d}t {\rm d}E$ is the emission spectrum rate of particle $i$ arising directly from Hawking radiation -- the so-called primary particles (hence the subscript p) -- and is given by~\cite{Hawking:1974sod,Hawking:1975zls}
\begin{equation}
\label{eq:primary}
    \frac{{\rm d}^2 N^i_{\rm p}}{{\rm d}t {\rm d}E}(E,M) = \frac{n^i_{\rm dof}\, \Gamma^i(M,E)}{2\pi (e^{E/T} \pm 1)}~,
\end{equation}
where $n^i_{\rm dof}$ is the number of degrees of freedom of particle $i$ (spin/helicity and color; see Table 3 in Appendix C of~\cite{Arbey:2019had}), $\Gamma^i(M,E)$ is the greybody factor/absorption coefficient of a wave packet scattering in the BH spacetime geometry off into infinity (where an observer would be located) and the $+(-)$ sign corresponds to fermions (bosons). Calculating the greybody factor $\Gamma^i$ is the nontrivial part of Eq.~(\ref{eq:primary}), as it depends not only on $M$ and $E$, but also implicitly on the radiated particle's mass, spin and any other internal degree of freedom it may possess. Although these greybody factors can be  approximated analytically~\cite{MacGibbon:1990duy,MacGibbon:1991tj}, we use the exact numerical calculation from the \verb+BlackHawk+ code \cite{Arbey:2019had}. For a given BH mass, \verb+BlackHawk+ calculates the instantaneous primary emissions ${\rm d}^2 N^i_{\rm p}/{\rm d}t {\rm d}E$ for every particle species of interest.

The observable fluxes of stable particles from an evaporating PBH are different from Eq.~(\ref{eq:primary}) because of hadronization, fragmentation and electroweak corrections, which lead to the emission of secondary particles. For example, the production of pions from hadronization of primary emitted quarks, and their subsequent decay, produces additional spectra of neutrinos and photons. Also, the electroweak corrections on any primary emission of species $i$ lead to secondary neutrino and photon spectra which become increasingly important at energies $\gtrsim1$~TeV (corresponding to the last $\sim 400$~s of the evaporating PBH). Typically, the secondary emissions have been either approximated semi-analytically, for example in~\cite{Page:1976gpe,Bugaev:2007py,Ukwatta:2016iba} for gamma ray yield and in~\cite{Halzen:1995skd} for neutrino yield, or calculated numerically by the \verb+PYTHIA+ code~\cite{Sjostrand:2015zea}, see for example~\cite{Baker:2021bhr} for gamma ray yield. Both methods have some degree of imprecision: in addition to approximated hadronization/fragmentation calculations, the electroweak corrections are usually absent in the semi-analytical method; in the \verb+PYTHIA+ code the electroweak corrections are not included completely (for example, triple gauge couplings are missing), which renders its range of reliability below tens of TeV. In order to improve on the secondary emission calculation, we use the \verb+HDMSpectra+ code~\cite{Bauer:2021vjs} which is valid up to Planck scale. Although \verb+HDMSpectra+ aims at computing the secondary emission from heavy dark matter annihilation/decay, it can be equally applied to evaporating PBHs; as soon as particle species $i$ is produced either from dark matter or PBH evaporation, the rest (secondary calculation) is the same in both cases. The secondary spectrum of particle $j$ can be obtained by the energy convolution of the primary spectrum of particle $i$, Eq.~(\ref{eq:primary}), with the spectrum of particle $j$ from a fixed energy of primary species $i$ (and summing over all species $i$):
\begin{equation}
\label{eq:secondary}
    \frac{{\rm d}^2 N^j_{\rm s}}{{\rm d}t {\rm d}E}(E,M) = \sum_{i} \int_0^\infty {\rm d}E_{\rm p}\, \frac{{\rm d}^2 N^i_{\rm p}}{{\rm d}t {\rm d}E_{\rm p}}(E_{\rm p},M) \frac{{\rm d} N^{i\rightarrow j}}{{\rm d}E}(E,E_{\rm p})~.
\end{equation}
In this relation, the ${\rm d} N^{i \rightarrow j}/{\rm d}E$ term has been computed by using \verb+HDMSpectra+. Some remarks about Eq.~(\ref{eq:secondary}) are in order: the summation index $i$ runs over particle and anti-particle of each species and the ${\rm d} N^{i \rightarrow j}/{\rm d}E$ should be computed for each one separately (for $u$, $\bar{u}$, $b$, $\bar{b}$, etc), which are provided via explicit fragmentation functions in \verb+HDMSpectra+ (using the \verb+HDMSpectra.FF+ syntax). In this case the $n^i_{\rm dof}$ in Eq.~(\ref{eq:primary}) includes the spin/helicity and color degrees of freedom. However, the rates of particle and anti-particle production by a PBH are equal. Thus, by doubling the $n^i_{\rm dof}$, Eq.~(\ref{eq:primary}) provides the production rate spectrum of particle plus anti-particle (for particles that are not identical to their anti-particle). By using this convention, one can use the \verb+HDMSpectra.spec+ syntax in \verb+HDMSpectra+ code which provides the secondary emission from $X\bar{X}$ ($X$ being the particle species). In this convention, the particle species $Y=\gamma, Z^0$, gluons and Higgs need a special treatment since the \verb+HDMSpectra.spec+ syntax provides the secondary emission from $YY$: we should divide their secondary spectra ${\rm d} N^{i \rightarrow j}/{\rm d}E$ by two. 

Considering both primary and secondary contributions, the \textit{total} emission rate spectrum of $\nu_\alpha$ ($\alpha = e,\mu,\tau$) and $\gamma$ from a PBH can be written as
\begin{equation}\label{eq:totnu}
    \frac{{\rm d}^2 N^{\nu_\alpha}_{\rm tot}}{{\rm d}t {\rm d}E} = \frac{{\rm d}^2 N^{\nu_\alpha}_{\rm s}}{{\rm d}t {\rm d}E} + \mathcal{A}^{\nu_\alpha\to \nu_\alpha}(E) \frac{{\rm d}^2 N^{\nu_\alpha}_{\rm p}}{{\rm d}t {\rm d}E}~,
\end{equation}
and
\begin{equation}\label{eq:totgamma}
    \frac{{\rm d}^2 N^{\gamma}_{\rm tot}}{{\rm d}t {\rm d}E} = \frac{{\rm d}^2 N^{\gamma}_{\rm s}}{{\rm d}t {\rm d}E} + \mathcal{A}^{\gamma\to \gamma}(E) \frac{{\rm d}^2 N^{\gamma}_{\rm p}}{{\rm d}t {\rm d}E} + \mathcal{A}^{Z^0\to \gamma}(E) \frac{{\rm d}^2 N^{Z^0}_{\rm p}}{{\rm d}t {\rm d}E}~,
\end{equation}
where $\mathcal{A}^{k\to j}(E)$ is the coefficient of $\delta(1-E/E_{\rm p})$ in the $k\to j$ secondary spectrum which takes into account the probability of no corrections for $\nu_\alpha\to\nu_\alpha$ and $\gamma\to\gamma$, and the probability of $\gamma$ production for $Z^0\to\gamma$. The $\mathcal{A}$ coefficients are provided by \verb+HDMSpectra+.

A remark on Eq.~(\ref{eq:totnu}) is in order: as can be inferred from this equation, we are assuming emission of neutrinos in flavor eigenstates via Hawking radiation and we suppose Majorana nature for neutrinos by setting $n_{\rm dof}^{(\nu_\alpha+\bar{\nu}_\alpha)}=2$. However, although it is undeniably more natural to suppose neutrino emission in mass eigenstates~\cite{Lunardini:2019zob}, since the majority of neutrino yield originates from secondary production ({\it i.e.}, small $\mathcal{A}^{\nu_\alpha\to \nu_\alpha}$), this assumption does not modify our conclusions by more than $\sim$~few percents. In the case of Dirac neutrinos, the number of degrees of freedom will be four~\cite{Lunardini:2019zob}, instead of two, which enhances the neutrino emission and facilitates the detection of PBHs (to be conservative, we continue with the Majorana neutrino assumption.) Finally, the nonzero neutrino masses (though $\lesssim0.01$~eV) would not affect the assumed massless-neutrino emission in \verb+BlackHawk+ code, for PBHs with $M\sim10^{15}$~g~\cite{Lunardini:2019zob}.

%%%%%%%%%            figure 1           %%%%%%%%%%%%%
%%%%%%%%%%%%%%%%%%%%%%%%%%%%%%%%%
\begin{figure}[t!]
\centering
\subfloat[]{
\includegraphics[width=0.49\textwidth]{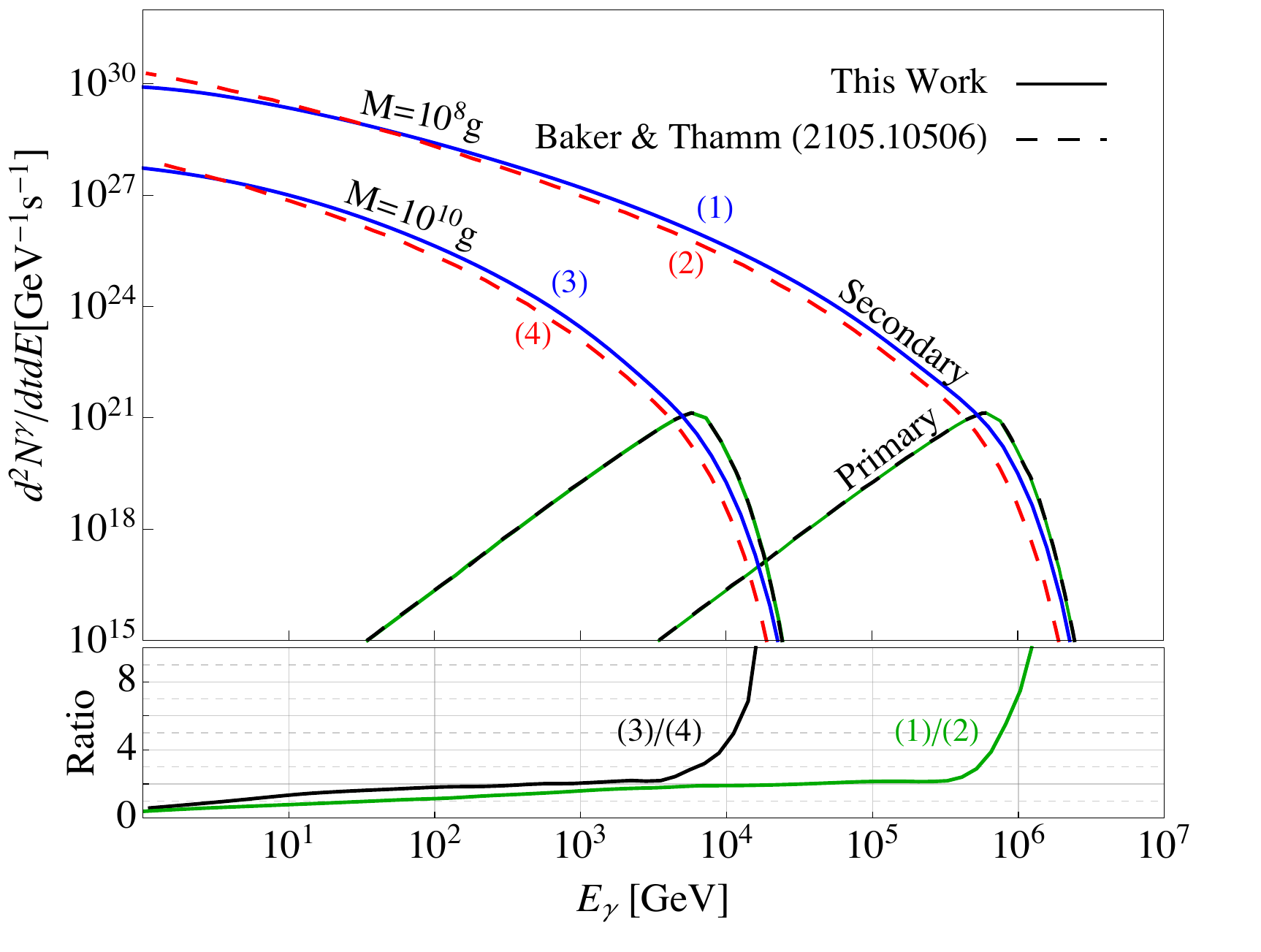}
\label{fig:gamma-a}
}
\subfloat[]{
\includegraphics[width=0.49\textwidth]{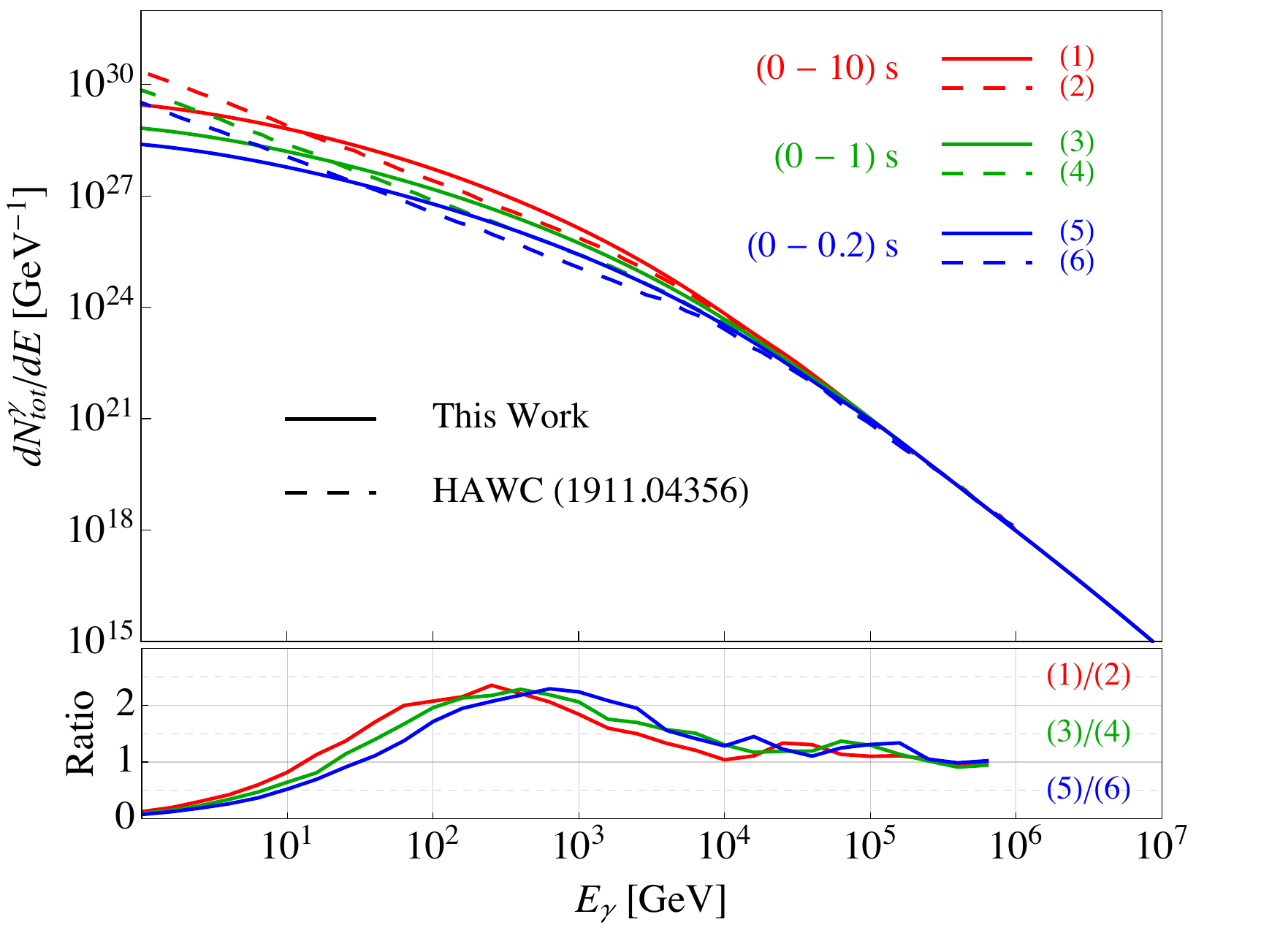}
\label{fig:gamma-b}
}
\caption{\label{fig:spec-gamma}(a) The instantaneous primary and secondary gamma ray spectra from an evaporating PBH. The solid curves show our results and the dashed curves are taken from~\cite{Baker:2021bhr}. The lower part shows the ratio of solid to dashed secondary emissions curves. (b) The time-integrated \textit{total} gamma ray spectra for three different time intervals in solid curves. The dashed curves show the spectra used by HAWC~\cite{Albert:2020doa}. The ratio of solid to dashed curves are depicted in the lower part.}
\end{figure}
%%%%%%%%%%%%%%%%%%%%%%%%%%%%%%%
%%%%%%%%%%%%%%%%%%%%%%%%%%%%%%%  

Figure~\ref{fig:spec-gamma} shows the gamma ray emission spectra. In the left panel the instantaneous spectra (both primary and secondary) are shown for two PBH masses $M=10^{10}$~g and $10^8$~g, corresponding respectively to $\sim 400$~s and $\sim 4\times10^{-4}$~s before the death. In order to compare with previous works, the primary and secondary gamma ray emission spectra from~\cite{Baker:2021bhr} are also shown by dashed curves; the ratios of our secondary emissions to the ones from~\cite{Baker:2021bhr} are depicted in the lower part of Figure~\ref{fig:gamma-a}. While the primary emissions are identical (both in this paper and in~\cite{Baker:2021bhr}, they are taken from the \verb+BlackHawk+ code), the ratio plot manifests a factor of $\sim2$ larger secondary gamma ray emission at $E_\gamma\gtrsim 100$~GeV in our computation. The ratio increases at higher energies for both PBH masses. The time-integrated \textit{total} spectra from Eq.~(\ref{eq:totgamma}) for three different time intervals are shown in Figure~\ref{fig:gamma-b}, in solid curves, with the corresponding time-integrated spectra used by the HAWC collaboration~\cite{Albert:2020doa} represented by dashed curves. The ratio plot at the bottom shows a factor $\sim2$ larger total spectra from our computation at $E_\gamma\sim$~TeV in all three time intervals. 

\begin{figure}[t!]
\centering
\subfloat[]{
\includegraphics[width=0.49\textwidth]{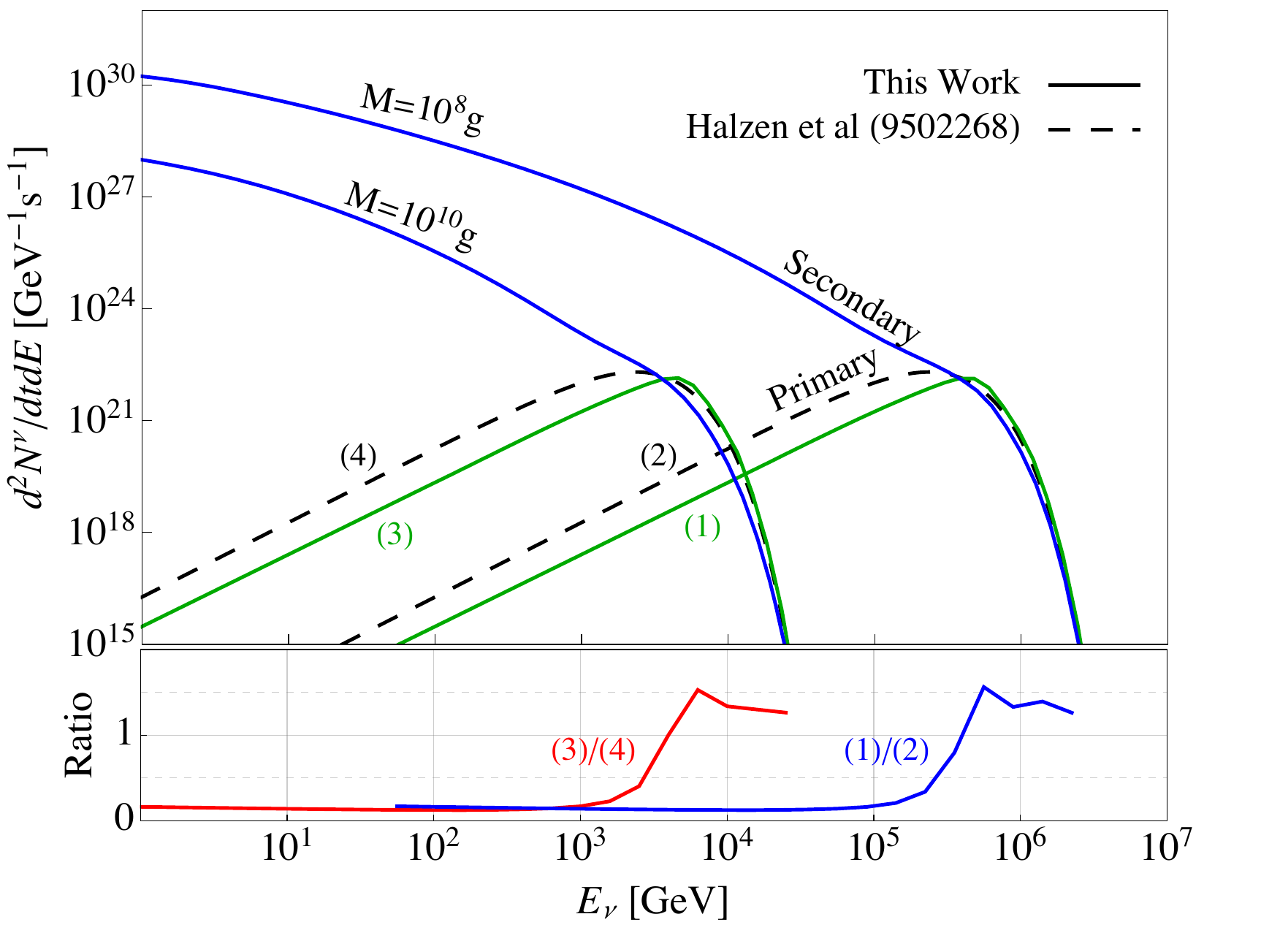}
\label{fig:nu-a}
}
\subfloat[]{
\includegraphics[width=0.49\textwidth]{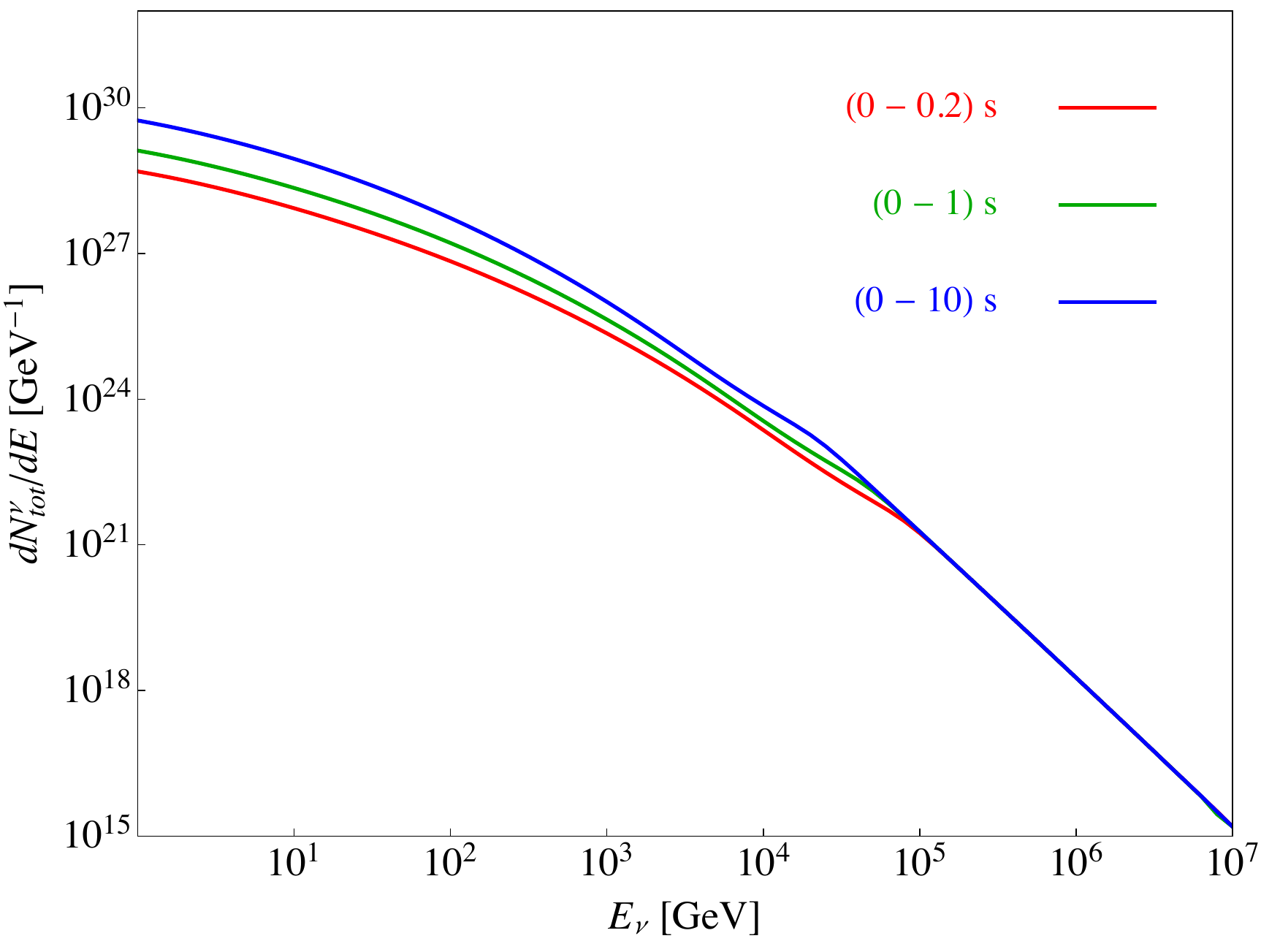}
\label{fig:nu-b}
}
\caption{\label{fig:spec-nu}The same as Figure~\ref{fig:spec-gamma}, but for neutrinos (sum of the three flavors of neutrinos and antineutrinos). The dashed curves in panel (a) show the primary neutrino emission of~\cite{Halzen:1995skd} which has been used by IceCube collaboration in~\cite{Dave:2019fus}. The ratio of solid to dashed curves of primary spectra are shown in the lower part of panel (a). Panel (b) shows the time-integrated total neutrino spectra for three different time intervals before the complete evaporation of PBH.}
\end{figure}

The expected neutrino spectra from evaporating PBHs are shown in Figure~\ref{fig:spec-nu}, where in both panels we plot the sum of all three flavors of neutrinos and antineutrinos. In the left panel, the primary and secondary instantaneous neutrino emissions for two PBH masses are depicted with solid curves and are compared with the primary neutrino emission of~\cite{Halzen:1995skd}, which has been used by IceCube collaboration in~\cite{Dave:2019fus}, in dashed curves. The lower ratio plot of Figure~\ref{fig:nu-a} illustrates a factor $\sim8$ smaller primary neutrino emission from \verb+BlackHawk+ with respect to the analytical calculation of~\cite{Halzen:1995skd} at energies below the peak of Hawking radiation. Figure~\ref{fig:nu-b} shows the time-integrated \textit{total} neutrino spectra from Eq.~(\ref{eq:totnu}), for three different time intervals, which we could not find any work in the literature to compare with.

%%%%%%%%%            figure 2           %%%%%%%%%%%%%
%%%%%%%%%%%%%%%%%%%%%%%%%%%%%%%%%
%%%%%%%%%%%%%%%%%%%%%%%%%%%%%%%%%%%%%
%%%%%%%%%%%%%%%%%%%%%%%%%%%%%%%%%%%%%

\begin{figure}[t!]
\centering
\subfloat[]{
\includegraphics[width=0.49\textwidth]{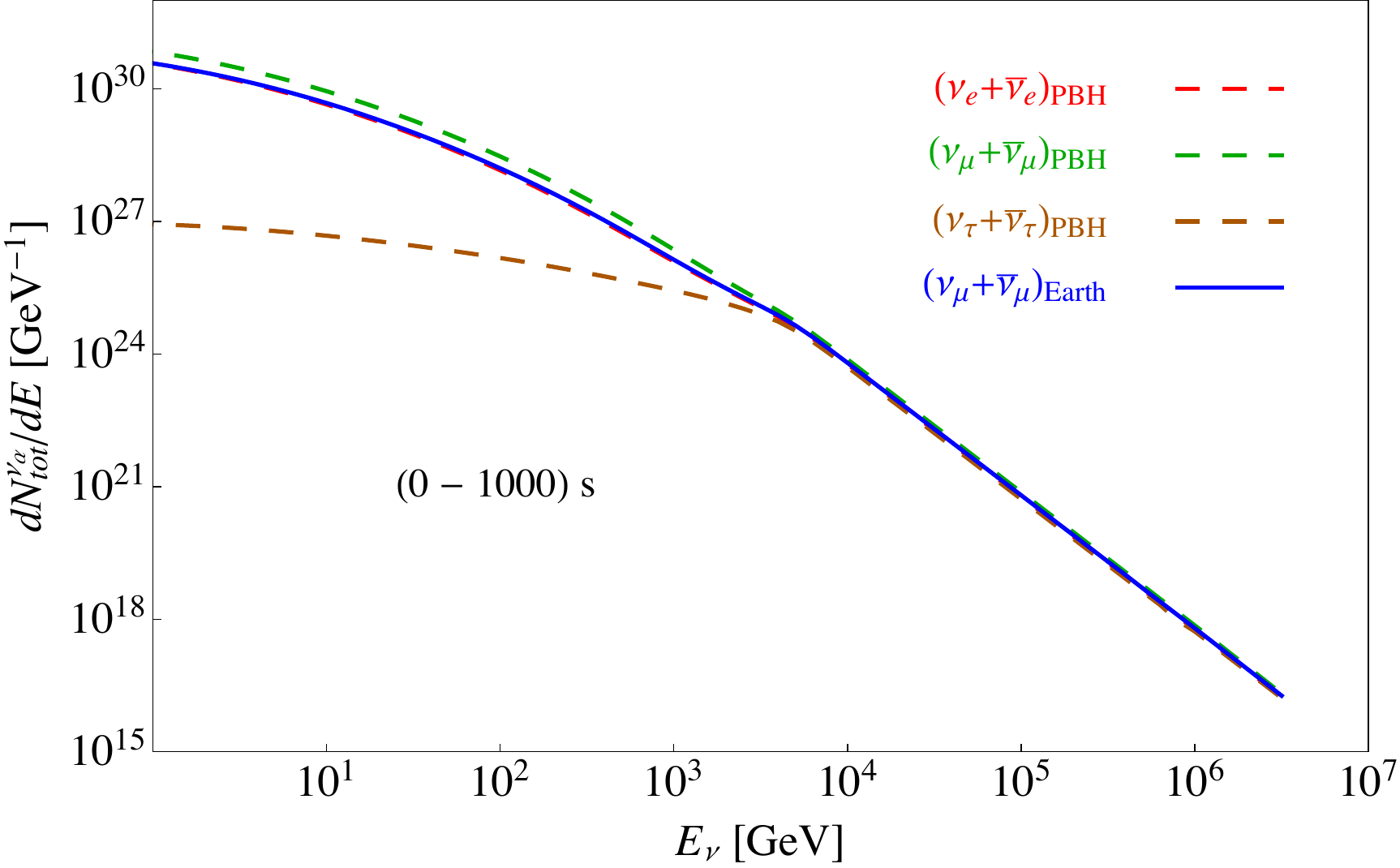}
\label{fig:osc-nu}
}
\subfloat[]{
\includegraphics[width=0.49\textwidth]{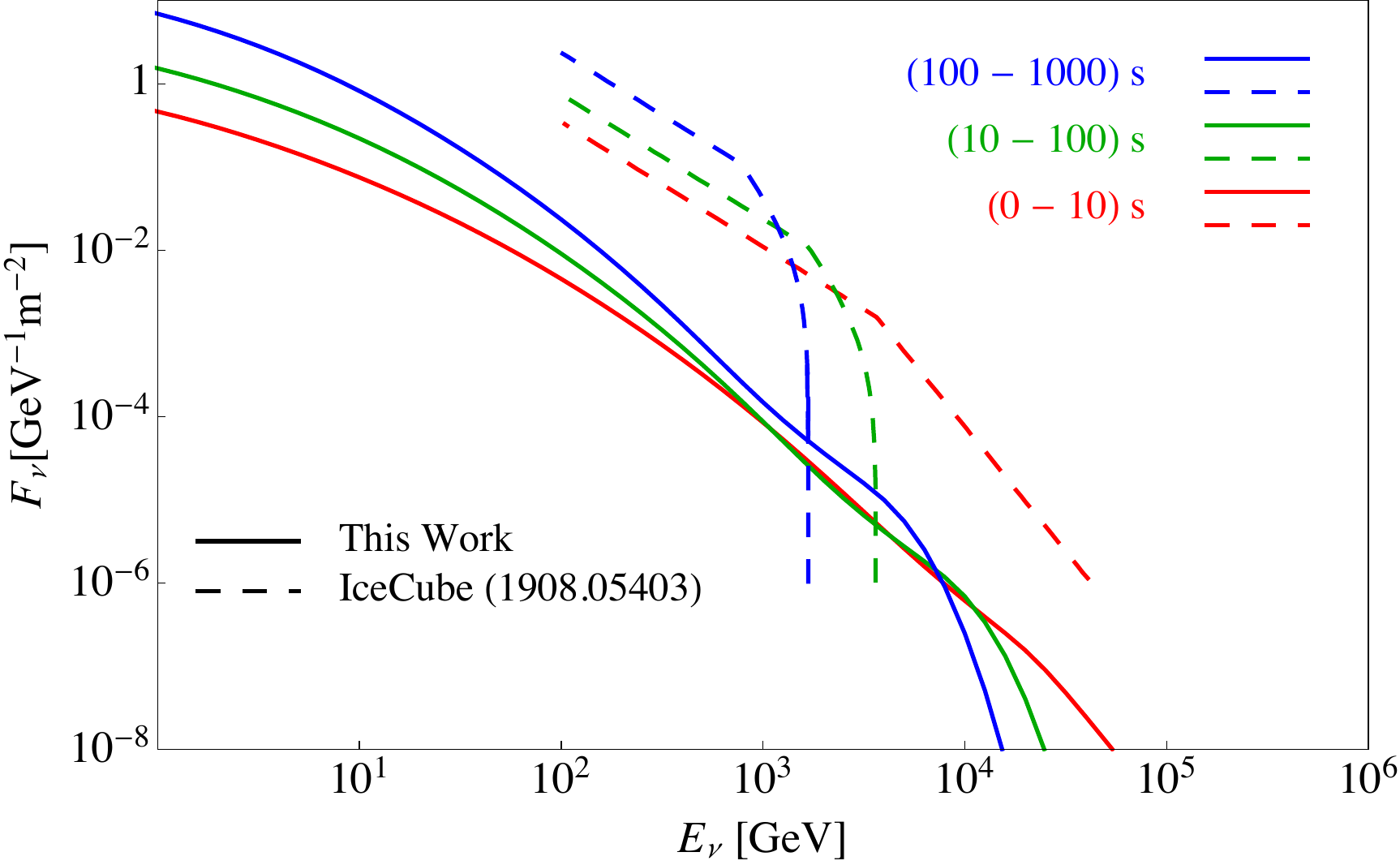}
\label{fig:flu}
}
\caption{\label{fig:3}(a) The total spectra of each neutrino flavor at the production site (PBH), in dashed curves, and the muon-flavor spectrum (neutrino and antineutrino) at the Earth in the solid curve, after taking into account the flavor oscillation in Eq.~(\ref{eq:osc}). (b) Solid curves: the time-integrated flux (fluence) of all neutrino flavors at the Earth for three different time intervals. Dashed curves: the fluences used in~\cite{Dave:2019fus}, which are based on the calculation of~\cite{Halzen:1995skd}. The distance to the PBH is $d_L=10^{-2}$~pc.}
\end{figure}

The spectrum of $\nu_\alpha$ at the Earth, where by $\nu_\alpha$ we mean the sum of neutrinos and antineutrinos of flavor $\alpha$, can be obtained by taking into account the decoherent flavor oscillations en route from PBH to Earth:
\begin{equation}\label{eq:osc}
    \frac{{\rm d}^2 N^{\nu_\alpha}_{\rm tot}}{{\rm d}t {\rm d}E}\bigg|_\oplus = \sum_{\beta = e}^\tau \sum_{i=1}^3 |U_{\alpha i}|^2 |U_{\beta i}|^2\; \frac{{\rm d}^2 N^{\nu_\beta}_{\rm tot}}{{\rm d}t {\rm d}E}\bigg|_{\rm PBH}~,
\end{equation}
where $U_{\alpha i}$ are the elements of the PMNS mixing matrix fixed to their best-fit values from the NuFit global analysis~\cite{Esteban:2020cvm}. The effect of oscillation can be visualized in Figure~\ref{fig:osc-nu}, which shows the prompt (at production) spectra of each neutrino flavor and the $\nu_\mu$ spectrum after oscillations. The time-integrated flux (fluence) of $\nu_\alpha$ from an evaporating PBH located at luminosity distance $d_L$ ($\approx d_c$ for small redshifts, where $d_c$ is the comoving distance) is given by 
\begin{equation}\label{eq:fluence}
    F_{\nu_\alpha}(E; t_i \rightarrow t_f) = \frac{1}{4\pi d_L^2} \int_{t_i}^{t_f} {\rm d}t\, \frac{{\rm d}^2 N^{\nu_\alpha}_{\rm tot}}{{\rm d}t {\rm d}E}\bigg|_\oplus~.
\end{equation}  
The same Eq.~(\ref{eq:fluence}) applies to the fluence of gamma rays, albeit without any flavor oscillation. The solid curves in Figure~\ref{fig:flu} show the fluence of all-flavor neutrinos at the Earth from a PBH at the distance $0.01$~pc for three different time intervals. For comparison, the fluences (all neutrino flavors) used in~\cite{Dave:2019fus} which are based on the calculation of~\cite{Halzen:1995skd} are displayed by dashed curves in Figure~\ref{fig:flu}. A clear mismatch, both qualitatively and quantitatively, can be seen in this figure.  

%%%%%%%%%            figure 3           %%%%%%%%%%%%%
%%%%%%%%%%%%%%%%%%%%%%%%%%%%%%%%%
%%%%%%%%%%%%%%%%%%%%%%%%%%%%%%%%%%%%%
%%%%%%%%%%%%%%%%%%%%%%%%%%%%%%%%%%%%%

\begin{figure}[t!]
\centering
\subfloat[]{
\includegraphics[width=0.49\textwidth]{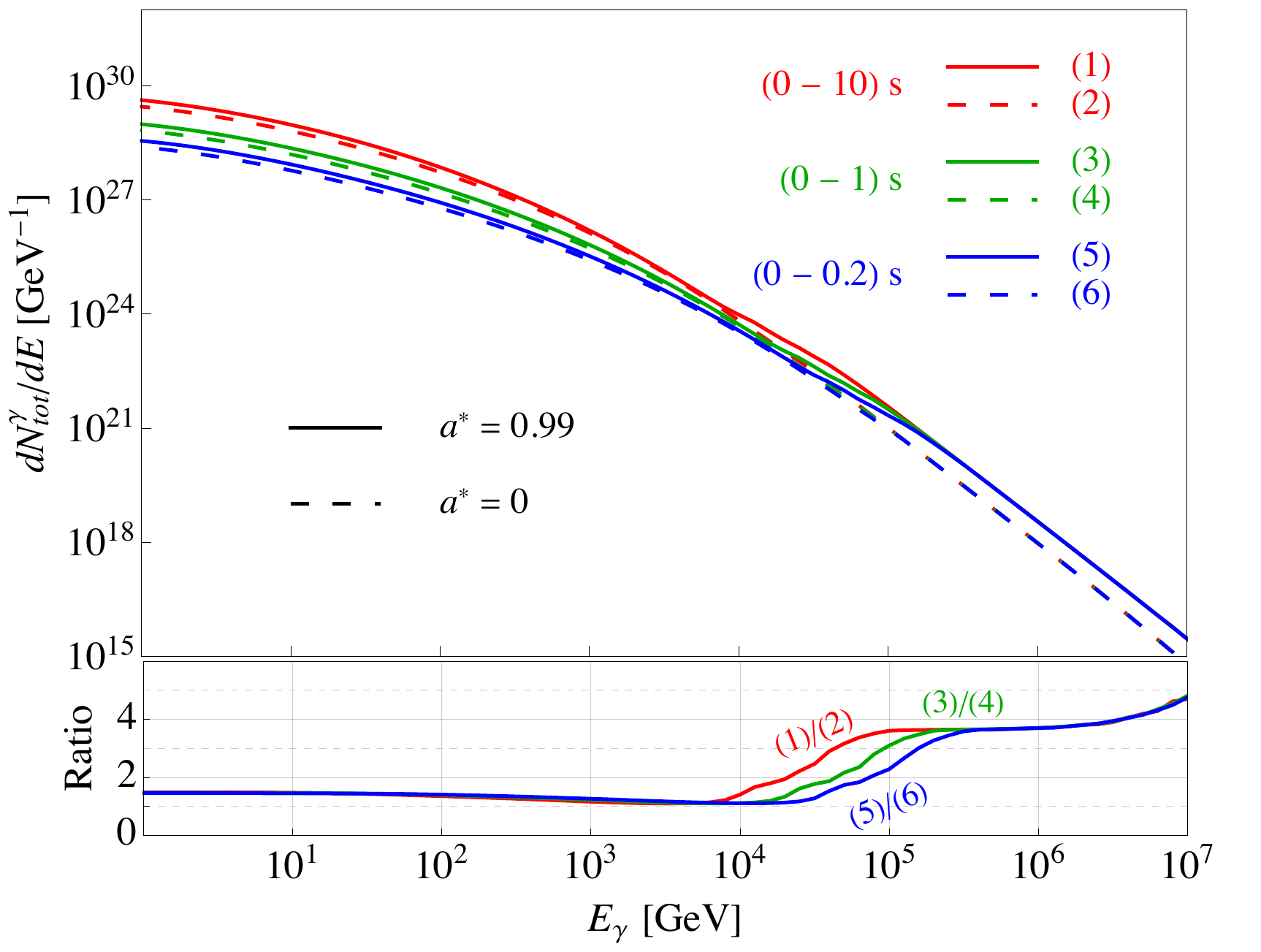}
\label{fig:spin-gamma}
}
\subfloat[]{
\includegraphics[width=0.49\textwidth]{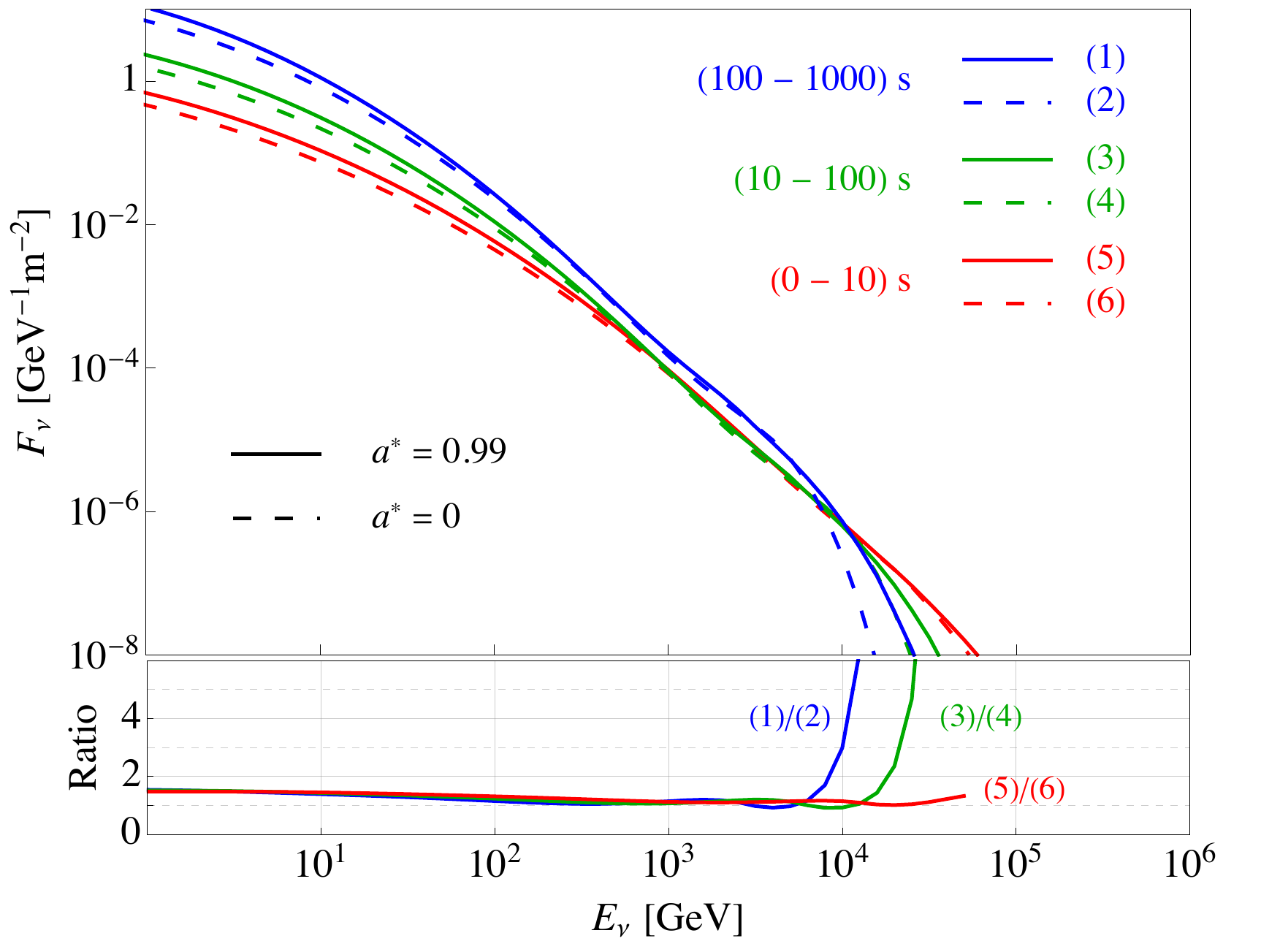}
\label{fig:spin-nu}
}
\caption{\label{fig:spin}The time-integrated gamma ray spectra (left panel) and fluence of neutrinos (right panel, assuming $d_L=10^{-2}$~pc) for Schwarzschild PBH ($a^\ast=J/M^2=0$) in dashed curves and maximally rotating PBH ($a^\ast=0.99$) in solid curves. The lower plots show the ratio of $a^\ast=0.99$ to $a^\ast=0$.}
\end{figure}

In the case of rotating PBHs, it is well-known that the neutrino and gamma ray (and all other particles) emissions will be amplified by the increase of angular momentum~\cite{Page:1976ycs}. As a consequence, the lifetime of PBHs shortens and effectively the rotating PBHs experience a faster runaway mass loss in the last stage of their life. Whether the nonzero angular momentum of a PBH helps its observation or not depends on the competition between the increase in emission spectra and decrease in the domain of time integration of spectra, or equivalently of the fluence in Eq.~(\ref{eq:fluence}), remembering that for rotating BHs a part of the energy budget is dedicated to rotation and this competition does not necessarily end in a draw. To illustrate the result, we choose the extreme case of a PBH with maximal spin $a^\ast=J/M^2=0.99$, where $J$ is the angular momentum, and assume that it keeps its angular momentum until the complete evaporation. The left and right panels of Figure~\ref{fig:spin} show, respectively, the time-integrated gamma ray spectra and fluence of neutrinos (assuming $d_L=10^{-2}$~pc) for $a^\ast=0\,(0.99)$ by dashed (solid) curves. The lower parts of panels depict the ratio of $a^\ast=0.99$ to $a^\ast=0$, which clearly show an increase in the time-integrated spectra of gamma rays and neutrinos for rotating PBHs. For neutrinos, especially in longer time intervals, the fluence is bigger at high energies for $a^\ast=0.99$, while at lower energies a rotating PBH produces marginally more neutrinos than a Schwarzschild PBH. Let us emphasize that the modifications in emission spectra depicted in Figure~\ref{fig:spin} are for PBHs which maximally rotate until the last moments of their life, a scenario that should be considered quite exotic since the evaporating PBHs are expected to be Schwarzschild BHs in the last stages of their life even if they had a large angular momentum at the early stages~\cite{Page:1976ycs,Arbey:2019jmj}. Therefore, although we do not consider further the rotating PBHs, it can be concluded from Figure~\ref{fig:spin} that possible residual tiny angular momenta of PBHs do not change the bounds that will be derived in the next sections for non-rotating PBHs; {\it i.e.}, these bounds are robust against possible small $a^\ast$ since even for maximal $a^\ast$ they tighten by a factor $\mathcal{O}(1)$ and not more. The limits on rotating PBHs will be provided in the next section.

%%%%%%%%%            figure 4           %%%%%%%%%%%%%
%%%%%%%%%%%%%%%%%%%%%%%%%%%%%%%%%
%%%%%%%%%%%%%%%%%%%%%%%%%%%%%%%%%%%%%
%%%%%%%%%%%%%%%%%%%%%%%%%%%%%%%%%%%%%

%%%%%%%%%%%%%%%%%%%%%%%%%%%%%%%%%
%%%%%%%%%%%%%%%%%%%%%%%%%%%%%%%%%
\section{\label{sec:detect}PBH detection prospects at IceCube}
%%%%%%%%%%%%%%%%%%%%%%%%%%%%%%%%%
%%%%%%%%%%%%%%%%%%%%%%%%%%%%%%%%%

To quantify the observability of evaporating PBHs as transient point sources in neutrino telescopes, let us calculate the expected number of events at IceCube from the fluence of Eq.~(\ref{eq:fluence}). We consider IceCube's $\mu$-track data set which benefits from a good angular resolution in the reconstruction of the incoming neutrino's direction ($\lesssim 1\degree$ above $\sim$~TeV energies). The good angular resolution assists in efficient rejection of background events from atmospheric neutrinos. The atmospheric muon background can be suppressed entirely by restricting to up-going $\mu$-track events in the detector, corresponding to the northern hemisphere.

\begin{figure}[t!]
\centering
\subfloat[]{
\includegraphics[width=0.49\textwidth]{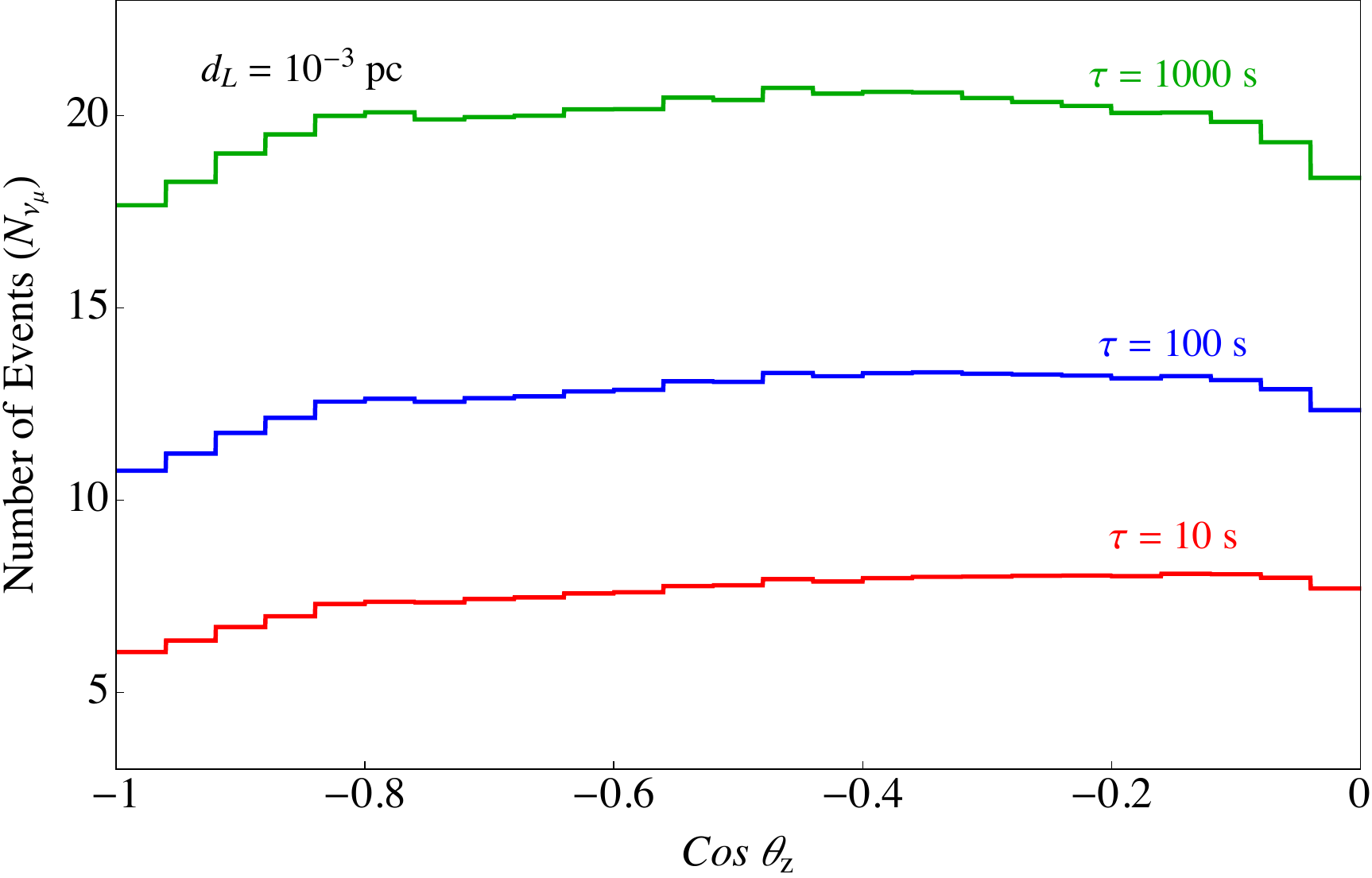}
\label{fig:event-numu}
}
\subfloat[]{
\includegraphics[width=0.49\textwidth]{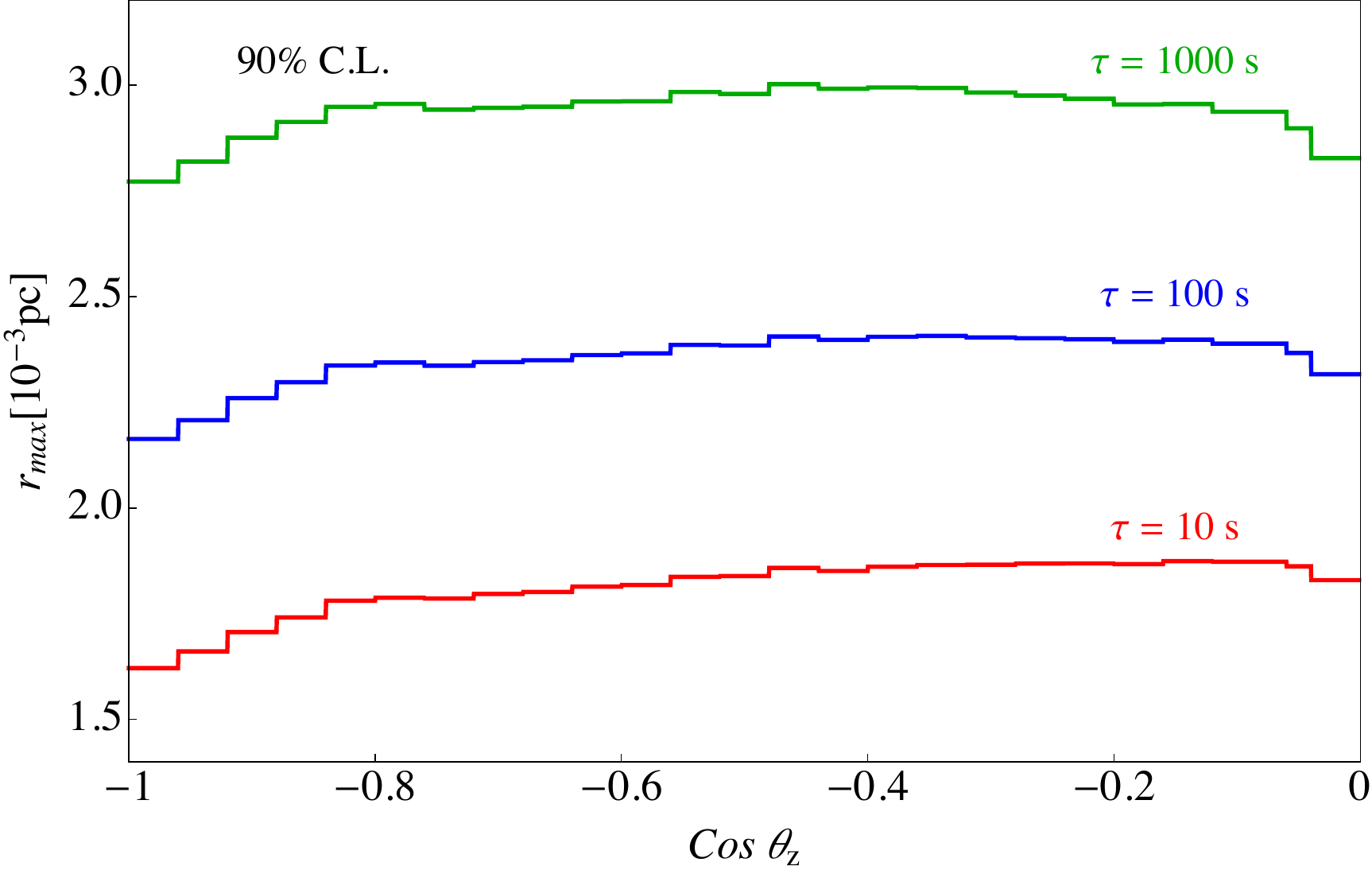}
\label{fig:rmax}
}
\caption{\label{fig:4}(a) The expected number of $\mu$-track events in IceCube from an evaporating PBH, located at the distance $d_L=10^{-3}$~pc and different zenith angles, in the last $\tau$ seconds before its death. (b) The maximum distance, $r_{\rm max}$, that can be probed by IceCube at 90\% C.L., for three time intervals $\tau$.}
\end{figure}

The expected number of $\mu$-track events from an evaporating PBH located at zenith angle $\theta_z$ and with fluence $F_{\nu_\mu}$ in the time interval $t_i\to t_f$ is given by
\begin{equation}\label{eq:nevents}
    N_{\nu_\mu}(\theta_z,t_i \to t_f) = \int_{E_{\rm min}}^{E_{\rm max}} {\rm d}E\, F_{\nu_\mu}(E; t_i \to t_f)\, A_{\rm eff}(E,\theta_z)~,
\end{equation}
where $A_{\rm eff}(E,\theta_z)$ is IceCube's effective area~\cite{Aartsen:2019fau}\footnote{In both neutrino telescopes and extensive air shower experiments, the azimuth-dependence of the effective area is very small and can be neglected.}. $E_{\rm min}$ is set by the energy threshold of $\mu$-track observation at IceCube, $\sim100$~GeV, and $E_{\rm max}$ is governed by the maximum energy of the fluence, which depends on the time interval $t_i\to t_f$. During the last $\sim10^3$~s of the life of a PBH at distance $\gtrsim10^{-3}$~pc from us, the proper motion due to PBH's motion is $\lesssim2^\circ$ (even for ultra-relativistic PBH velocities). The Earth's rotation also leads to $\lesssim0.5^\circ$ change in the position of PBH. Since the effective area of IceCube changes smoothly as a function of zenith angle, we can assume a fixed angular position for the PBH. Figure~\ref{fig:event-numu} shows the expected number of $\mu$-track events from an evaporating PBH located at distance $d_L=10^{-3}$~pc and at different zenith angles of the northern hemisphere, in the last $\tau$ seconds of its life. Clearly, as can be evidenced from Figure~\ref{fig:flu}, a larger time interval leads to a larger number of events at IceCube. However, any further increase of the time interval beyond $\sim10^3$~s results in a marginal gain since the lower temperature of the PBH implies a lower energy neutrino emission, which eventually falls below IceCube's threshold.

%%%%%%%%%            figure 5           %%%%%%%%%%%%%
%%%%%%%%%%%%%%%%%%%%%%%%%%%%%%%%%
%%%%%%%%%%%%%%%%%%%%%%%%%%%%%%%%%%%%%
%%%%%%%%%%%%%%%%%%%%%%%%%%%%%%%%%%%%%

The sensitivity of IceCube to a PBH depends on the number of atmospheric neutrino background events. Assuming a fixed-position PBH and considering $1\degree$ uncertainty in the direction reconstruction, the expected number of atmospheric neutrino events in the time interval of $\tau=10$~s is $\lesssim 10^{-5}$ (we used the HKKM atmospheric neutrino flux~\cite{Honda:2015feo}). Since the atmospheric neutrino flux is practically constant in time, the expected number of background events scales with $\tau$. Thus, as a consequence of the short transient nature of a PBH signal, the observation is background free\footnote{Strictly speaking, one has to search for clustering of events in a fixed time interval sliding over the data-taking time and properly incorporate the expected number of background events over the whole observation time. However, since performing this search requires detailed information of the detector which is not available, we opt for the simpler approach explained in the text.}. To estimate the constraint on the local density of PBHs, we use the Bayesian upper limit $N_{\nu_\mu}<N_{\rm max}=2.3$ at $90\%$ C.L. from the Poisson distribution, for our zero background expectation (see~\cite{Esmaili:2012us} for the details). From Eqs.~(\ref{eq:fluence}) and (\ref{eq:nevents}), this condition corresponds to the maximum distance
\begin{equation}
    r_{\rm max}(\theta_z; \tau) = \sqrt{\frac{1}{4\pi N_{\rm max}} \int_{E_{\rm min}}^{E_{\rm max}} {\rm d}E\, A_{\rm eff}(E,\theta_z) \int_{0}^{\tau} {\rm d}t\, \frac{{\rm d}^2 N_{\rm tot}^{\nu_\mu}}{{\rm d}t {\rm d}E}\bigg|_\oplus}~,
\end{equation}
beyond which PBH signal events are compatible with the background. Summing over all the zenith bins in the northern hemisphere (the $i$th bin is defined by $\theta_{z,{\rm min}}^i\leq\theta_z^i\leq\theta_{z,{\rm max}}^i$), the volume that can be probed by IceCube is
\begin{equation}\label{eq:maxV}
    V_{\rm max}(\tau) = \frac{1}{3} \sum_{i} \Omega_i r_{\rm max}^3(\theta_z^i; \tau)~,
\end{equation}
where $\Omega_i = 2\pi(\cos\theta_{z,{\rm min}}^i - \cos\theta_{z,{\rm max}}^i)$. Figure~\ref{fig:rmax} shows $r_{\rm max}$ as a function of zenith angle for time interval searches $\tau = 10,10^2$ and $10^3$~s before the PBH's death. The upper limit on local rate density of PBH bursts $\Dot{\rho}_{\rm max}$ (in [pc$^{-3}$yr$^{-1}$]) can be obtained via
\begin{equation}
    \Dot{\rho}_{\rm max}(\tau) = \frac{N_{\rm max}}{V_{\rm max}(\tau) T}~,
\end{equation}
where $T$ is the total observation time ($\sim 10$ years for IceCube). Table \ref{tab:limits} reports the limits on $\Dot{\rho}_{\rm max}$ found in this work for three different time intervals. Of course, these limits are not competitive with the current best constraints from HAWC, 3400~pc$^{-3}$ yr$^{-1}$ at $99\%$~C.L.~\cite{Albert:2020doa}, yet they show the capability of current neutrino telescopes in the search for PBHs. The factor of $\gtrsim 10^3$ difference is mostly due to $\sim 2$ -- $4$ orders of magnitude smaller effective area of IceCube with respect to HAWC's~\cite{Abeysekara:2013ecn}. For rotating PBHs with maximal angular momentum ($a^\ast=0.99$), the limits reported in Table \ref{tab:limits} are tighter by $\sim (15-30)$\% due to their slightly larger fluences in Figure \ref{fig:spin-nu} (in combination with the energy-dependence of IceCube's effective area which peaks at $\sim100$~TeV.) The advantages of using neutrino telescopes in the search for PBHs include the larger field of view with respect to gamma ray experiments (in case of one rare event that can be located outside the directional acceptance of these experiments), situations in which gamma rays will be absorbed before reaching the Earth, and simultaneous observation of PBHs in neutrino and gamma ray experiments which eases its identification (the latter will be discussed in the next section). Finally, it should be mentioned that the limit on $\Dot{\rho}_{\rm max}$ does not improve significantly for searches in timescales longer than $\sim10^3$~s.

%%%%%%%%%%%%   Table 1 %%%%%%%%%%%%%%%%
%%%%%%%%%%%%%%%%%%%%%%%%%%%%%%%%%%%%%%%
\begin{table}[t]
\centering
\caption{\label{tab:limits}Limits, at 90\% C.L., on the local rate density of PBH bursts, $\Dot{\rho}_{\rm max}$, from the ten-years IceCube $\mu$-track data set, for three different time intervals before the PBH's death, for both non-rotating and rotating PBHs.}
\vspace{0.4cm}
\begin{tabular}{|c|c|c|}
\hline
$\tau$ {[}s{]} & $\Dot{\rho}_{\rm max}(a^*=0)$ {[}pc$^{-3}$ yr$^{-1}${]} & $\Dot{\rho}_{\rm max}(a^*=0.99)$ {[}pc$^{-3}$ yr$^{-1}${]} \\ 
\hline
$10$ & $1.7 \times 10^7$ & $1.5 \times 10^7$ \\ 
\hline
$10^2$ & $8.2\times10^6$ & $6.2 \times 10^6$ \\ 
\hline
$10^3$ & $4.1\times10^6$ & $3.3 \times 10^6$ \\                              
\hline
\end{tabular}
\end{table}
%%%%%%%%%%%%%%%%%%%%%%%%%%%%%%%%%%%%%%%
%%%%%%%%%%%%%%%%%%%%%%%%%%%%%%%%%%%%%%%

%%%%%%%%%%%%%%%%%%%
\section{\label{sec:gamma-nu}PBH in gamma ray experiments and multimessenger correlations}
%%%%%%%%%%%%%%%%%%%

Before discussing the multimessenger observation of an evaporating PBH, let us comment on the HAWC limit on $\Dot{\rho}_{\rm max}$. The number of gamma ray events at HAWC can be calculated analogously to $N_{\nu_\mu}$ in Eq.~(\ref{eq:nevents}) by substituting the effective area $A_{\rm eff}$ and the fluence accordingly. Our re-evaluation of the gamma ray emission and its ratio to the emission spectra used by HAWC collaboration, shown in Figure~\ref{fig:gamma-b}, reveals quantitative differences. Since the majority of event counts in HAWC are in the energy range $\lesssim10$~TeV, where the ratio plot of Figure~\ref{fig:gamma-b} points to a factor $\sim2$ larger number of events in our evaluation, it corresponds to a factor $\sim1.4$ larger $r_{\rm max}$ that can be probed by HAWC. By taking into account this raise, the current HAWC limit on the local PBH rate density ($3400$~pc$^{-3}$ yr$^{-1}$) can be improved to roughly $\Dot{\rho}_{\rm max} \lesssim 1200$~pc$^{-3}$ yr$^{-1}$ at $99\%$~C.L.

\begin{figure}[t!]
\centering
\subfloat[]{
\includegraphics[width=0.49\textwidth]{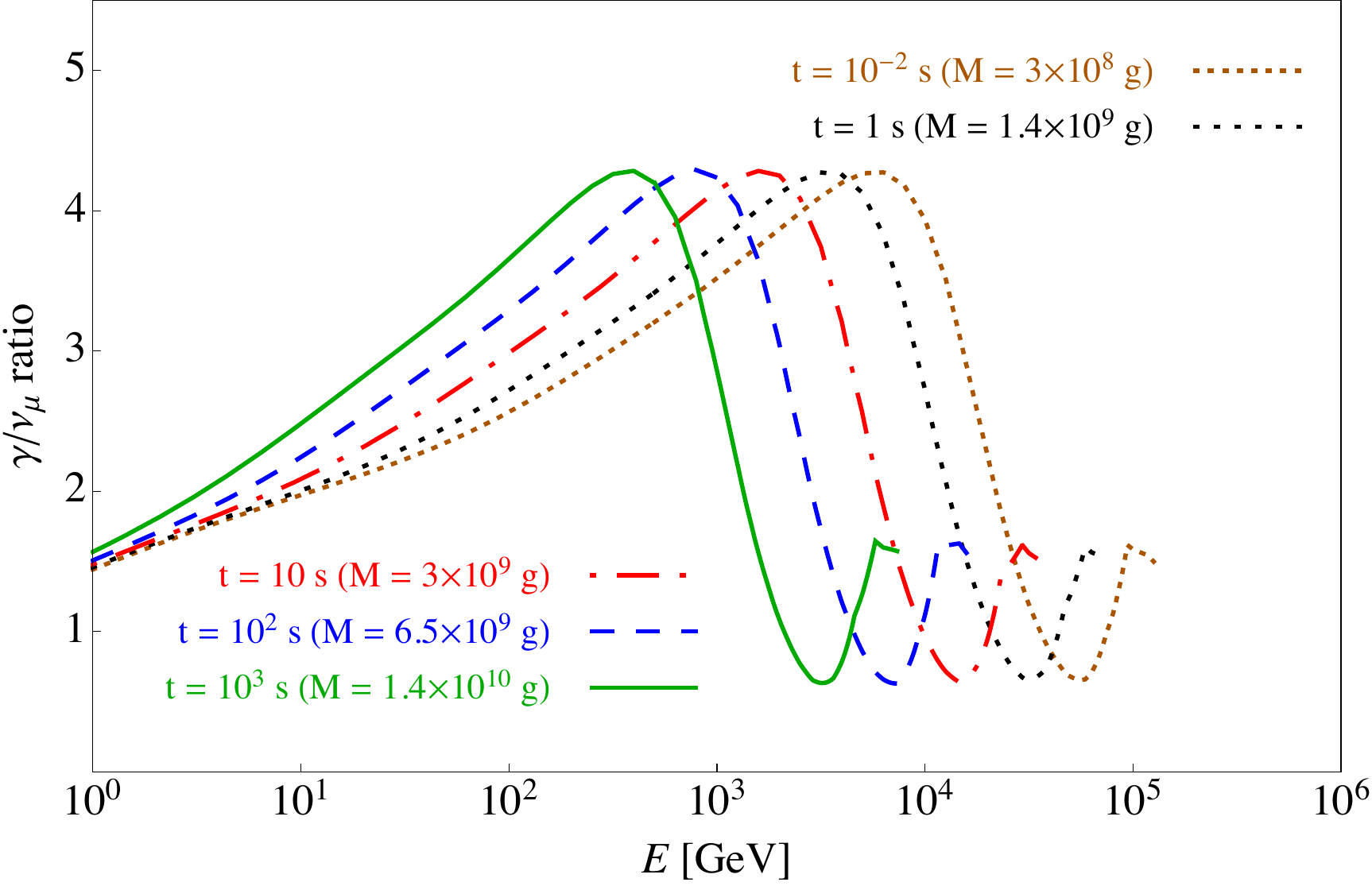}
\label{fig:ratio}
}
\subfloat[]{
\includegraphics[width=0.49\textwidth]{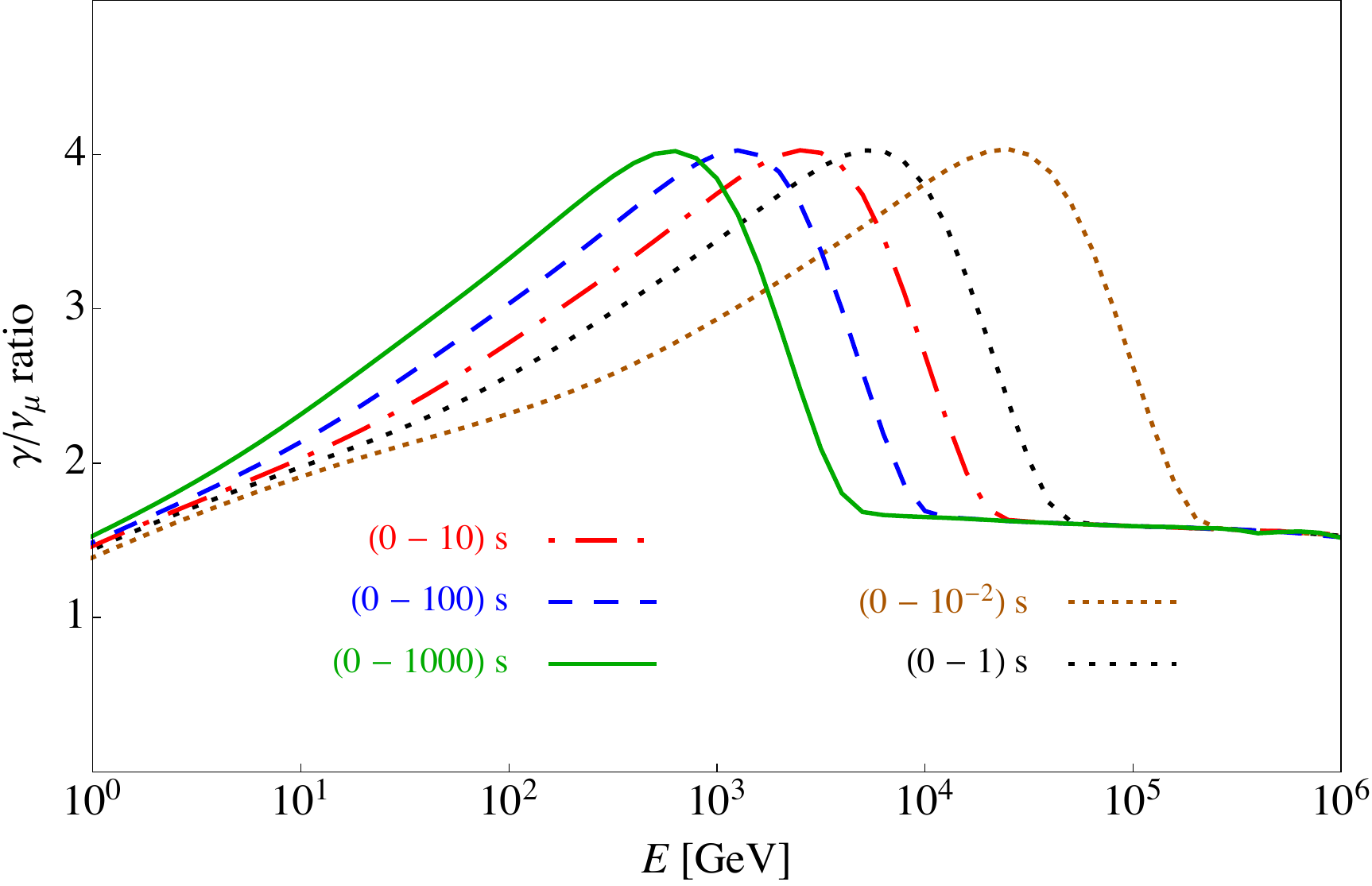}
\label{fig:r-energy}
}
\caption{\label{fig:5}(a) Ratio of the instantaneous $\gamma/\nu_\mu$ spectra at various time instants. (b) Ratio of time-integrated spectra of gamma rays and neutrinos for several time intervals.}
\end{figure}

Observing any astrophysical event with more than one messenger is always beneficial. While the gamma ray experiments offer the best chance to detect a burst from an evaporating PBH, it might be hard to distinguish the observed signal from other transient gamma ray sources, or pinpointing it unambiguously to a PBH. Unlike the conventional $pp$ and $p\gamma$ scenarios for steady-state emitting sources, the relation between gamma ray and neutrino emission spectra for transient (explosive) sources can be nontrivial and changes depending on the source conditions. However, the exact ratio between the photon and neutrino spectra (at different times and energies) contains a unique signature of an evaporating PBH. Figure~\ref{fig:ratio} shows the energy-dependence of $({\rm d}^2 N^{\gamma}_{\rm tot}/{\rm d}t {\rm d}E)/({\rm d}^2 N^{\nu_\mu}_{\rm tot}/{\rm d}t {\rm d}E|_\oplus)$ at $t =10^{-2}, 1,10,10^2$ and $10^3$~s before the death of PBH. The peculiar, and almost $t$-independent, features of the curves in Figure~\ref{fig:ratio} are the consequence of both the thermal Hawking radiation and the secondary production through hadronization and electroweak corrections in Standard Model. At the highest energies (close to the corresponding energy for each $t$) the $\sim1.5$ gamma to neutrino ratio results from the greybody factors in Eq.~(\ref{eq:primary}) and the $\mathcal{A}$ coefficients of Eqs.~(\ref{eq:totnu}) and (\ref{eq:totgamma}). The drop in the ratio just below the highest energy, to $\sim0.5$, and the rise in lower energies, to $\sim4$, are the consequence of the secondary production mechanisms: while both the hadronization and electroweak corrections participate in photon and neutrino secondary productions, the former is more efficient in photon production at lower energies and the latter predominantly creates neutrinos with slightly lower energies than the primary energy. This energy separation in photon and neutrino secondary productions is an outcome of the well-established Standard Model physics, and in harmony with the quasi-black-body radiation from PBHs, shapes the drop and rise of the curves in Figure~\ref{fig:ratio}, carrying the signature of PBHs.  

%%%%%%%%%            figure 6           %%%%%%%%%%%%%
%%%%%%%%%%%%%%%%%%%%%%%%%%%%%%%%%
%%%%%%%%%%%%%%%%%%%%%%%%%%%%%%%%%%%%%
%%%%%%%%%%%%%%%%%%%%%%%%%%%%%%%%%%%%%

\begin{figure}[t!]
\centering
\subfloat[]{
\includegraphics[width=0.49\textwidth]{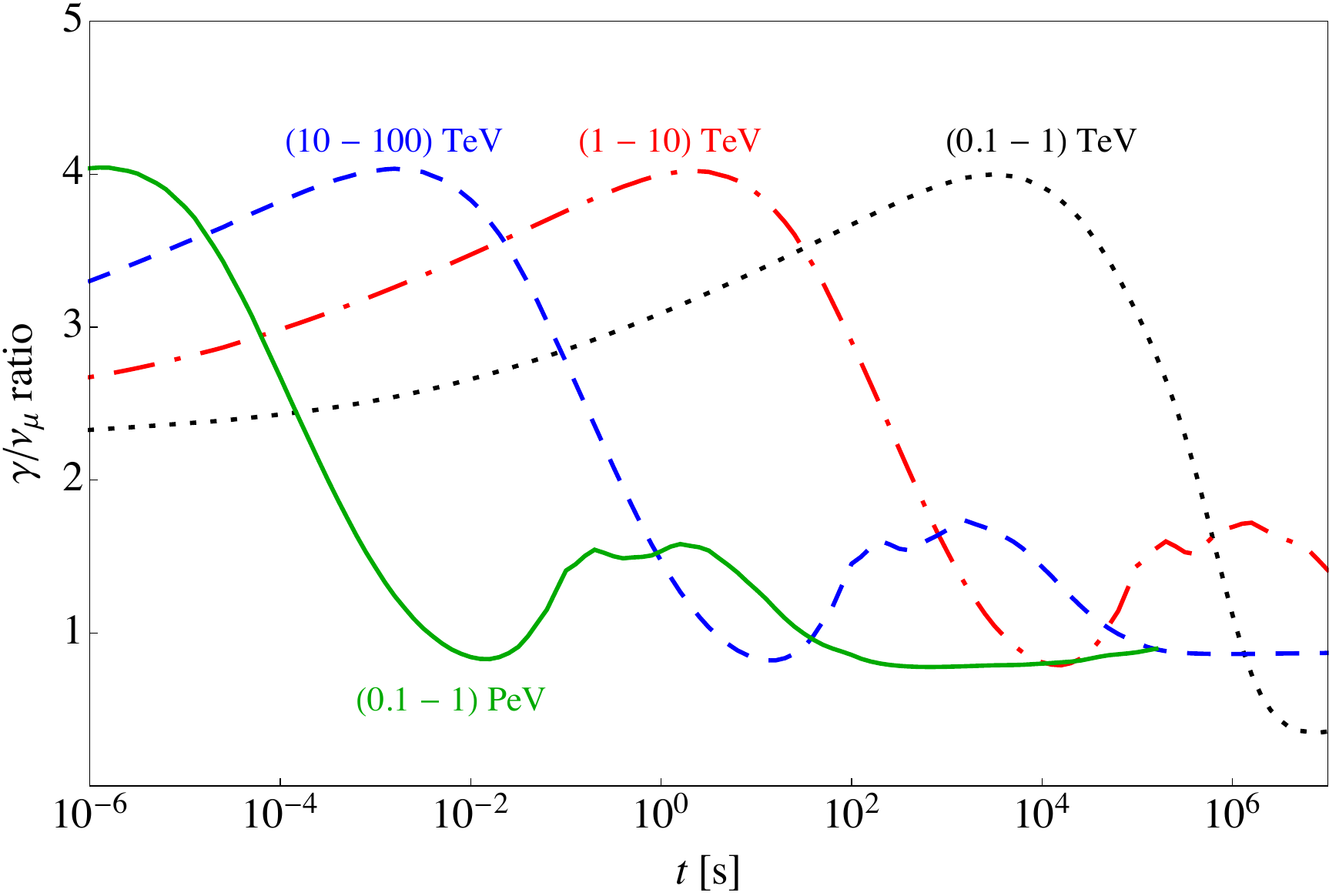}
\label{fig:r-time}
}
\subfloat[]{
\includegraphics[width=0.51\textwidth]{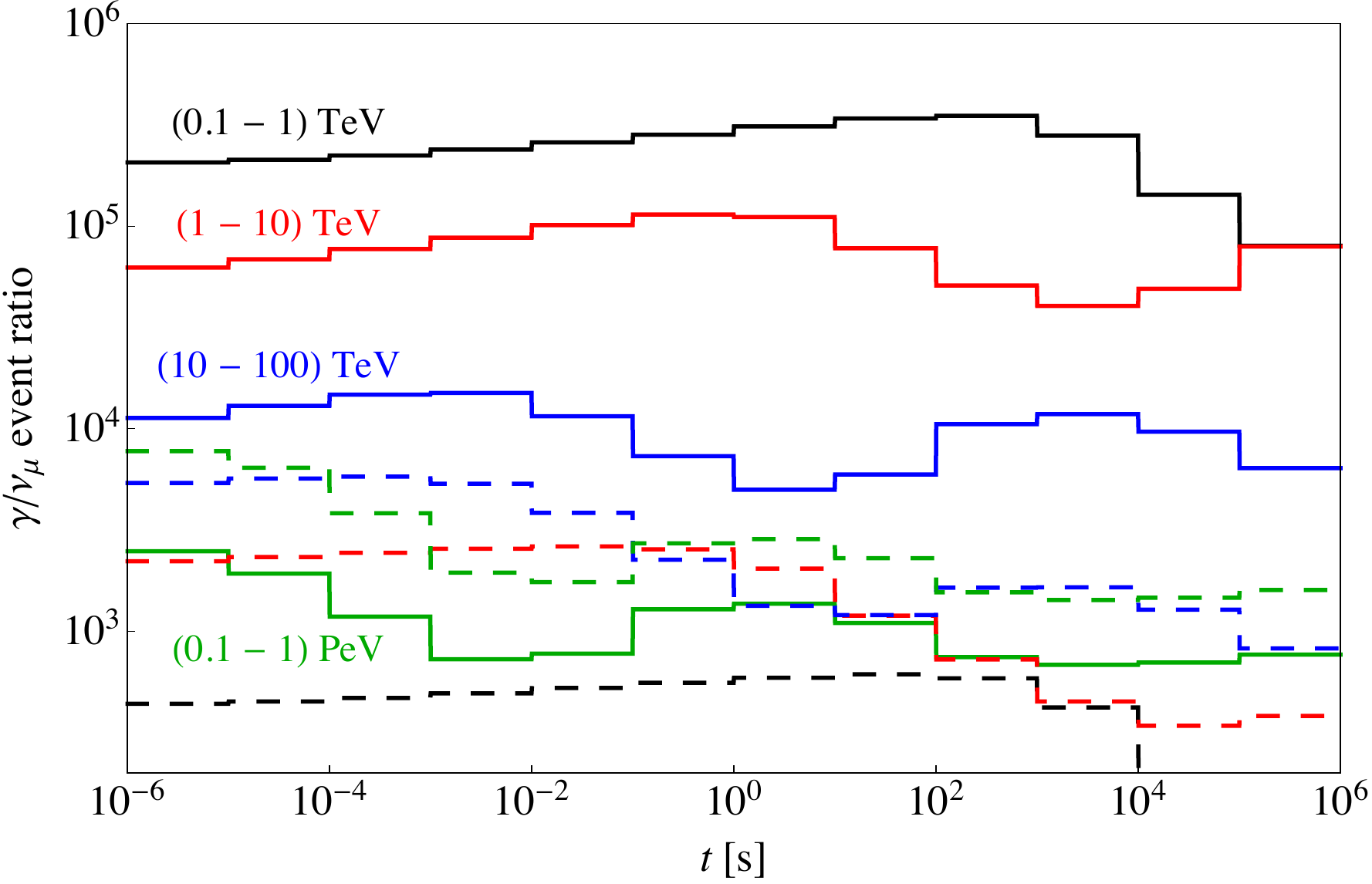}
\label{fig:r-events}
}
\caption{\label{fig:6}(a) Ratio of the energy-integrated spectra of gamma rays to neutrinos in the stated bins of energy. (b) Ratio of the number of gamma ray to neutrino events, respectively at HAWC and IceCube experiments, for a PBH located at declination $\delta = 20^\circ (70^\circ)$ in solid (dashed) curves.} 
\end{figure}

Although the features of the curves in Figure~\ref{fig:ratio} seem quite compelling, from an observational point of view, which requires some integration over energy and/or time, these features would change.  Figure~\ref{fig:r-energy} shows the ratio of time-integrated spectra, $$\left. \int_0^\tau({\rm d}^2 N^{\gamma}_{\rm tot}/{\rm d}t {\rm d}E){\rm d}t\middle/\int_0^\tau({\rm d}^2 N^{\nu_\mu}_{\rm tot}/{\rm d}t {\rm d}E|_\oplus){\rm d}t\right.~,$$  
as function of energy and for the time intervals $\tau=10^{-2}, 1,10,10^2$ and $10^3$~s. Time integration leads to the disappearance of the dips in Figure~\ref{fig:ratio}, while the energy-dependent rise of the gamma ray spectrum with respect to the neutrino's is still evident in Figure~\ref{fig:r-energy}. For larger $\tau$, the peak of $\gamma/\nu_\mu$ ratio moves to lower energies due to the lower PBH temperature. The energy-integrated spectra of gamma rays and neutrinos are in fact more important since both the extensive air shower experiments and neutrino telescopes have limited energy resolutions, while the time resolution is favorable (at the $\sim$ nanosecond level). Figure~\ref{fig:r-time} shows the ratio of the energy-integrated spectra (over the energy bins indicated in the figure) of gamma rays to neutrinos as function of the remaining time to the complete evaporation of PBH. In lower (higher) energy bins, the peak of $\gamma/\nu_\mu$ ratio is located at larger (smaller) times, although the gamma ray spectrum keeps dominating at $t$ values below the peak. These features are the direct consequence of the dips in Figure~\ref{fig:ratio}.

All the features and the correlations between gamma ray and neutrino spectra illustrated in Figures~\ref{fig:5} and \ref{fig:r-time} can be verified and be used to infer the PBH nature of the observed event in both the extensive air shower and neutrino detectors. At the level of number of gamma ray and neutrino events, and the correlation among them, detector characteristics such as respective effective areas, the location of PBH and experimental cuts should be taken into account. To illustrate how the ratio of gamma ray to neutrino events looks like, in Figure~\ref{fig:r-events} we show this ratio as function of time and integrated over energy bins for two different locations of PBH: the solid (dashed) curves correspond to a PBH locating at declination $\delta = 20^\circ (70^\circ)$, where the HAWC has its maximum (minimum) sensitivity, while for both cases IceCube's sensitivity remains almost the same. The drops and rises of the curves in Figure~\ref{fig:r-time} are somehow visible also in Figure~\ref{fig:r-energy}, though modified by the effective areas (compare, for example, the red and blue curves in both panels). We have to emphasize that both the gamma ray and neutrino experiments can infer the observed flux from their number of events distributions and therefore verify the features in Figures~\ref{fig:5} and \ref{fig:r-time}; of course when the PBH is located close enough to produce statistically adequate number of events, that is $d_L\lesssim10^{-4}$~pc, and within the field of view of both experiments.   

%%%%%%%%%            figure 7           %%%%%%%%%%%%%
%%%%%%%%%%%%%%%%%%%%%%%%%%%%%%%%%
%%%%%%%%%%%%%%%%%%%%%%%%%%%%%%%%%%%%%
%%%%%%%%%%%%%%%%%%%%%%%%%%%%%%%%%%%%%

%%%%%%%%%%%%%%%%%%%
\section{\label{sec:concl}Discussions and conclusions}
%%%%%%%%%%%%%%%%%%%

While the existence of PBHs has not yet been confirmed experimentally, the search for these remnants of primordial density fluctuations continues as a powerful diagnostic in probing early universe physics and probably the nature of dark matter. In this paper, we have focused on PBHs with initial mass $M\sim 10^{15}$~g, which are expected to be currently in their last stages of evaporation due to Hawking radiation and hence producing detectable bursts of stable high-energy particles, including gamma rays and neutrinos, appearing as transient point sources in observatories. By employing the state-of-the-art \verb+BlackHawk+ and \verb+HDMSpectra+ codes, we have re-evaluated the primary and secondary gamma ray and neutrino emissions, pointing out important quantitative and qualitative differences from the existing literature. For the gamma rays, the new expected flux leads to a factor $\sim3$ improvement on the local PBH burst rate density limit reported by HAWC collaboration, setting it to $\Dot{\rho}_{\rm max} \lesssim 1200$~pc$^{-3}$ yr$^{-1}$ at $99\%$~C.L.

For the neutrinos, by using the new flux evaluation which is significantly different from the few existing approximate calculations, we have estimated the IceCube's limit on PBH rate density from the 10-years $\mu$-track data set to $\Dot{\rho}_{\rm max} \lesssim 4\times10^6$~pc$^{-3}$ yr$^{-1}$ at $90\%$~C.L. Although, as was expected, neutrino detectors cannot compete with gamma rays experiments in constraining PBH rate density, the larger field of view of neutrino telescopes (compared to the limited field of view of gamma ray experiments) renders them useful in PBH searches, especially with the upcoming KM3Net~\cite{Adrian-Martinez:2016fdl} which guarantees $4\pi$~sr coverage of the sky. Unfortunately, future extensions of IceCube detector both in the high-energy, IceCube-Gen2~\cite{Aartsen:2014njl}, and low energy, IceCube Upgrade~\cite{Ishihara:2019cep}, does not boost the PBH search. Even though IceCube-Gen2's effective area at $\gtrsim$~PeV energies is expected to increase by a factor of $\sim10$, the PBH's neutrino fluence at $\gtrsim$~PeV is several orders of magnitude lower than that around $\sim$~TeV energies, and no gain can be envisaged. In the low energy side ($\lesssim100$~GeV) the increase in the neutrino emission from PBH is not as stiff as the atmospheric neutrino flux which is the main target of IceCube Upgrade.

Finally, we have shown how a multimessenger approach, taking into account the correlation between gamma ray and neutrino spectra, can provide a rapid identification of a PBH with distance $\lesssim10^{-4}$~pc, in the case of a simultaneous observation in gamma ray and neutrino experiments. The primary thermal radiation of PBHs and the well-established secondary emission due to hadronization and electroweak corrections predict peculiar time and energy profiles of gamma-to-neutrino ratio which can be used in distinguishing PBHs from other astrophysical transients and has been discussed thoroughly. The last argument, one more time, brings forward the importance of multimessenger approach in the era of precision measurements in astrophysics, that we are entering, and the crucial role of neutrinos in this field.

%\medskip
\begin{acknowledgments}
We thank Pasquale D. Serpico for reading the manuscript and his valuable comments. We thank comments by Yuber F. Perez-Gonzalez and J\'er\'emy Auffinger. The authors thank John David Rogers Computing Center (CCJDR) in the Institute of Physics “Gleb Wataghin”, University of Campinas, for providing computing resources. A. C. thanks the support received by the CNPq scholarship No. 140316/2021-3 and by the CAPES/PROEX scholarship No. 88887.511843/2020-00.
A.F. E. thanks the support received by the CAPES/PROEX scholarship No. 88887.617120/2021-00.
\end{acknowledgments}

\bibliographystyle{JHEP}
\bibliography{refs}

\end{document}